\documentclass[]{aastex631}

\newcommand{\B}[1]{{\color{blue}{#1}}}
\received{August 8, 2023}
\revised{August 22, 2023}
\accepted{August 24, 2023}

\shorttitle{Deciphering Slow-rise Precursor}
\shortauthors{Cheng et al.}
\graphicspath{{./}{figures/}}

\begin{document}

\title{Deciphering The Slow-rise Precursor of a Major Coronal Mass Ejection}

\author[0000-0003-2837-7136]{X. Cheng}
\affiliation{School of Astronomy and Space Science, Nanjing University, Nanjing 210093, China; \B{xincheng@nju.edu.cn}\\}
\affiliation{Max Planck Institute for Solar System Research, G$\ddot{o}$ttingen 37077, Germany\\}

\author{C. Xing}
\affiliation{School of Astronomy and Space Science, Nanjing University, Nanjing 210093, China; \B{xincheng@nju.edu.cn}\\}
\affiliation{Sorbonne Universit\'e, Observatoire de Paris - PSL, \'Ecole Polytechnique, Institut Polytechnique de Paris, CNRS, Laboratoire de physique des plasmas (LPP), 4 place Jussieu, F-75005 Paris, France\\}

\author{G. Aulanier}
\affiliation{Sorbonne Universit\'e, Observatoire de Paris - PSL, \'Ecole Polytechnique, Institut Polytechnique de Paris, CNRS, Laboratoire de physique des plasmas (LPP), 4 place Jussieu, F-75005 Paris, France\\}
\affiliation{Rosseland Centre for Solar Physics, Institute for Theoretical Astrophysics, Universitetet i Oslo, P.O. Box 1029, Blindern, 0315 Oslo, Norway\\}

\author{S. K. Solanki}
\affiliation{Max Planck Institute for Solar System Research, G$\ddot{o}$ttingen 37077, Germany\\}

\author{H. Peter}
\affiliation{Max Planck Institute for Solar System Research, G$\ddot{o}$ttingen 37077, Germany\\}

\author{M. D. Ding}
\affiliation{School of Astronomy and Space Science, Nanjing University, Nanjing 210093, China; \B{xincheng@nju.edu.cn}\\}

\begin{abstract}
Coronal mass ejections (CMEs) are explosive plasma phenomena prevalently occurring on the Sun and probably on other magnetically active stars. However, how their pre-eruptive configuration evolves toward the main explosion remains elusive. Here, based on comprehensive observations of a long-duration precursor in an event on 2012 March 13, we determine that the heating and slow rise of the pre-eruptive hot magnetic flux rope (MFR) are achieved through a precursor reconnection located above cusp-shaped high-temperature precursor loops. It is observed that the hot MFR threads are built up continually with their middle initially showing an ``M" shape and then being separated from the cusp of precursor loops, causing the slow rise of the entire MFR. The slow rise in combination with thermal-dominated hard X-ray source concentrated at the top of the precursor loops shows that the precursor reconnection is much weaker than the flare reconnection of the main eruption. We also perform a three-dimensional magnetohydrodynamics simulation that reproduces the early evolution of the MFR transiting from the slow to fast rise. It is also disclosed that it is the magnetic tension force pertinent to ``M"-shaped threads that drives the slow rise, which, however, evolves into a magnetic pressure gradient dominated regime responsible for the rapid-acceleration eruption.
\end{abstract}


\section{Introduction}
Stellar mass ejections and accompanying flaring result in energetic events \citep{maehara12,argiroffi19} that may prevent life from thriving on orbiting exo-planets \citep{dong18}. At present, these issues can best be studied in the solar system thanks to the direct visibility of coronal mass ejections (CMEs) and solar flares that release a large quantity of magnetized plasma ($\sim$10$^{11}$--10$^{13}$ kg), strong electromagnetic radiation and sub-relativistic energetic particles into the heliosphere \citep{forbes06,chen11_review,schmieder15}. When directed toward the Earth, CMEs interact with the magnetosphere and ionosphere and can disrupt communications, overload power grids, and present a hazard to astronauts \citep{gosling93,webb00,solanki04}.

The energetic eruptions are essentially consequences of the destabilization and reconfiguration of the coronal magnetic field. Prior to such eruptions, in a long-lasting quasi-static phase, the magnetic field, in particular above the polarity inversion line (PIL) of active regions (ARs), is gradually stressed by various flows at the photosphere, resulting in accumulation of magnetic free energy \citep{Cheung14}. The stressed magnetic fields are organised in an orderly fashion as either sheared arcades or a magnetic flux rope (MFR, a coherent structure with all field lines wrapping around its central axis). Regardless of the difficulty in directly measuring the coronal magnetic field, some observables serve as proxies of pre-eruptive magnetic configurations including filaments/prominences \citep{kuperus74,mackay10,schmieder13}, sigmoids \citep{hudson98,rust96,green07}, cavities \citep{gibson06_ssr,wangym10} and hot channels (coherent plasma structure with a temperature above 8 MK \citep{zhang12,cheng13_driver}). Among them, the hot channels and analogues seem to be a promising proxy of the MFR, which can even be used for prediction, as they usually appear prior to the eruption \citep{zhang12}, sometimes for hours \citep{patsourakos13,Nindos20}, and then continuously evolve toward the eruptions \citep{cheng13_driver,gou19,mitra_2019apj}. 

Nevertheless, how these pre-eruptive configurations evolve, in particular, toward the very onset of the fast eruption is yet to be ascertained \citep{aulanier14_reviewer,aulanier21_na}. In order to initiate the eruption, many physical mechanisms have been proposed including tether-cutting and breakout reconnection \citep{moore01,antiochos99} and ideal MHD instabilities etc. \citep{forbes91,torok04,kliem06}. Although evidence has been presented for the action of individual process \citep{moore01,williams05,chenhd14,cheng20}, it could be extremely difficult for a sole mechanism to initiate a real eruption. Many comprehensive observational studies suggest that the initiation process of CMEs is much more complicated than expected, multiple physical processes are often coupled to each other even though the dominated one may change from one phase to the other \citep{cheng20}. Once the eruption has been initiated, the dynamic energy release via runaway magnetic reconnection is switched on, during which the different structural components of CMEs are quickly formed and accelerated, giving rise to flare radiation at the same time \citep{priest02,linjun15,veronig2018}.

The other important but still puzzling characteristic during the early rise phase is that the pre-eruptive MFR is found to be much, almost one order of magnitude, hotter than the background quiet corona of 1-3 MK \citep{cheng12_dem}, which is true for over half of major eruptions based on a statistical survey \citep{nindos15}. It is speculated that the heating is most likely due to magnetic reconnection \citep{dudik14}. One piece of evidence is that the pre-eruptive hot MFR shows an increase in toroidal flux and stays stable for hours before it erupts successfully \citep{patsourakos13}. On the other hand, a number of pre-flare activities are detected prior to the eruption such as slow rise of pre-eruptive configurations \citep[e.g.,][]{zhang01,sterling05,kliem14,McCauley15,cheng20}, H$\alpha$ line broadening of pre-eruptive filaments \citep[e.g.,][]{cho2016SoPh}, enhancement of soft X-ray emission \citep[e.g.,][]{zhang06,priest14}, brightenings at the footpoints of chromospheric kernels \citep[e.g.,][]{wanghaimin17,chenhc_2019}, changes in magnetic topology \citep[e.g.,][]{chintzoglou15,liulijuan18} and even appearance of non-thermal particles \citep[e.g.,][]{syntelis16,Awasthi18,HernandezPerez19}. All these pre-flare activities are also suggested to be more or less caused by magnetic reconnection. However, justifying a clear and integrated physical picture that links the formation, heating and early rise of the MFR, as well as various observed pre-flare characteristics, to magnetic reconnection remains rather difficult. The major difficulty is that these pre-eruptive signatures are often of short-lived ($\sim$minutes), in particular for those from ARs \citep{cheng20}. Moreover, it is also limited by observational capacity, e.g., the field-of-view of instruments (such as Goode Solar Telescope) is too small to observe the entire pre-eruptive structure that usually approximates to the size of ARs, and/or only the signatures of pre-flare activities in the lower atmosphere were detected \citep{wanghaimin17}. Furthermore, it is almost hardly to disentangle critical characteristics in real observations as which are more complex than models. Such a problem becomes even worse due to inevitable projection effects \citep{zhang17,zhou_17,gou19,Awasthi18,HernandezPerez19}.

Here, through comprehensive analyses of a long-duration precursor phase of a major CME/flare on 2012 March 13 which overcomes part of aforementioned limitations, we dislcose the relations intertwined among the heating and early rise of the pre-eruptive hot MFR, various pre-flare characteristics and precursor reconnection. With a combination of observation-inspired three-dimensional magnetohydrodynamics simulation, it is suggested that the magnetic tension force within the MFR drives the slow rise; while the magnetic pressure gradient one is responsible for the following acceleration eruption.

\section{Instruments and Data}
The Atmospheric Imaging Assembly \citep[AIA;][]{lemen12} on board Solar Dynamics Observatory \citep[SDO;][]{pesnell12} images the solar atmosphere almost simultaneously with ten passbands covering temperatures from 0.06 to 20 MK. The temporal cadence and spatial resolution of seven (two) EUV (UV) passbands are 12 (24) s and 1.2 arcseconds, respectively. Among the seven EUV passbands, the 131 {\AA} and 94 {\AA}, sensitive to the high-temperature plasma above 6 MK, are used for detecting the hot pre-eruptive MFR; the 171 {\AA}, 193 {\AA}, 211 {\AA} and 335 {\AA} for observing the large-scale background corona; the 304  {\AA}, 1600 {\AA} and 1700 {\AA} for searching for signals of the eruption in the lower atmosphere \citep{odwyer10}. The EUV imaging data from the Sun Earth Connection Coronal and Heliospheric Investigation \citep[SECCHI;][]{howard08} on board the Solar Terrestrial Relations Observatory (STEREO), which was separated from SDO by an angle of $\sim$110$^{\circ}$ at the time the analysed observations were made, provide the second perspective on the eruption even though with a lower cadence (5 minutes) and resolution (3.2 arcseconds). 

In order to locate the reconnection during the precursor phase and reveal its physical properties, we utilize the Ramaty High Energy Solar Spectroscopic Imager \citep[RHESSI;][]{lin02} that is capable of performing X-ray imaging and spectroscopic diagnostics for hot plasma and accelerated electrons. The hard X-ray images are reconstructed with the Clean algorithm based on the detectors 3, 5, 6, 7, 8 and 9. The X-ray spectra from the detector 3 are analysed in detail to derive the temperature of hot plasma and low energy cut-off of accelerated electrons. The spectra from the other detectors are also examined but not showed here because the results from all detectors were very similar. 

In addition, to inspect the CME generated by the MFR eruption, we also make use of the data from the Large Angle and Spectrometric Coronagraph \citep[LASCO;][]{brueckner95} on board the Solar and Heliospheric Observatory (SOHO) and the SECCHI instrument suite on board STEREO-A. The CME velocity in the higher corona is from the CDAW Data Center\footnote{https://cdaw.gsfc.nasa.gov/}. The Geostationary Operational Environmental Satellite (GOES) provides the soft X-ray (SXR) fluxes of associated flares at two bands of 1-8 {\AA} and 0.5-4 {\AA}. The Helioseismic and Magnetic Imager \citep[HMI;][]{schou12} also on board SDO provides the photospheric vector magnetic field of the investigated AR with a temporal cadence of 12 minutes and spatial resolution of 1.2 arcsecond.

\section{Results}

\begin{figure*} 
      \centerline{\hspace*{-0.05\textwidth}
      \includegraphics[width=0.7\textwidth,clip=]{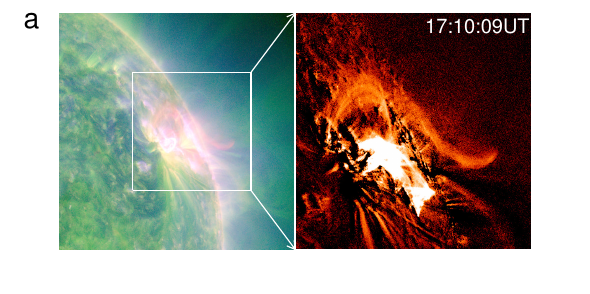}}\vspace*{-0.04\textwidth}
      \centerline{\hspace*{0.02\textwidth}
      \includegraphics[width=0.8\textwidth,clip=]{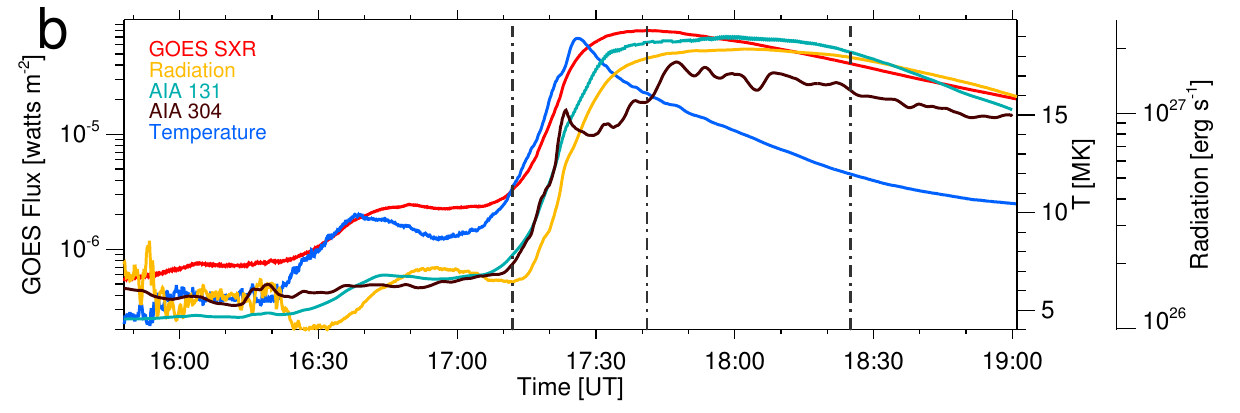}}\vspace*{-0.02\textwidth}
\caption{\textbf{Pre-eruptive MFR and pre-flare emission.} a. A composite of the AIA 131~{\AA} (red) and 171~{\AA} (green) images showing the pre-eruptive hot MFR and induced precursor loops (left) for the eruption on 2012 March 13. The AIA 131~{\AA} difference image (subtracting the image on 15:50~UT) displaying zoom-in of the MFR (right). b. Temporal evolution of the GOES SXR 1--8~{\AA} flux (red), temperature (blue), X-ray energy loss rate (yellow), integrated AIA 131~{\AA} (cyan) and 304~{\AA} (brown) intensity for NOAA AR 11429 with the FOV shown by the white box in panel a. The three dash-dotted lines indicate the onset, peak and end time of the main flare phase, respectively.} \label{f_flare}
\end{figure*}

\subsection{Overview of major CME eruption}
On 2012 March 13, an M7.9 class flare, taking place in the NOAA AR 11429 (Figure \ref{f_flare}a), started at $\sim$17:12~UT, peaked at $\sim$17:41~UT and ended at $\sim$18:25~UT (Figure \ref{f_flare}b). It also produced prominent 30 THz emissions \citep{Kaufmann13,trottet_2015} and was accompanied by a very energetic CME that included an erupting MFR as its main body. The projected average speed of the CME was over 1900 km s$^{-1}$, as measured in the field of view of LASCO. Figure \ref{f_flare}a shows that a loop-like pre-eruptive structure had appeared prior to the main flare and was associated with a long-lasting cusp-shaped precursor structure. Using the visibility of the cusp-shaped structure, such a precursor is found to last for more than one hour, much longer than that usually observed (several minutes) in previous events \citep{zhang12,cheng13_driver,HernandezPerez19}, thus enabling us to decipher the heating and early rise of the pre-eruptive hot MFR.

\subsection{Formation and heating of pre-eruptive MFR}

\begin{figure*}
      \centerline{\hspace*{-0.05\textwidth}
      \includegraphics[width=1.0\textwidth,clip=]{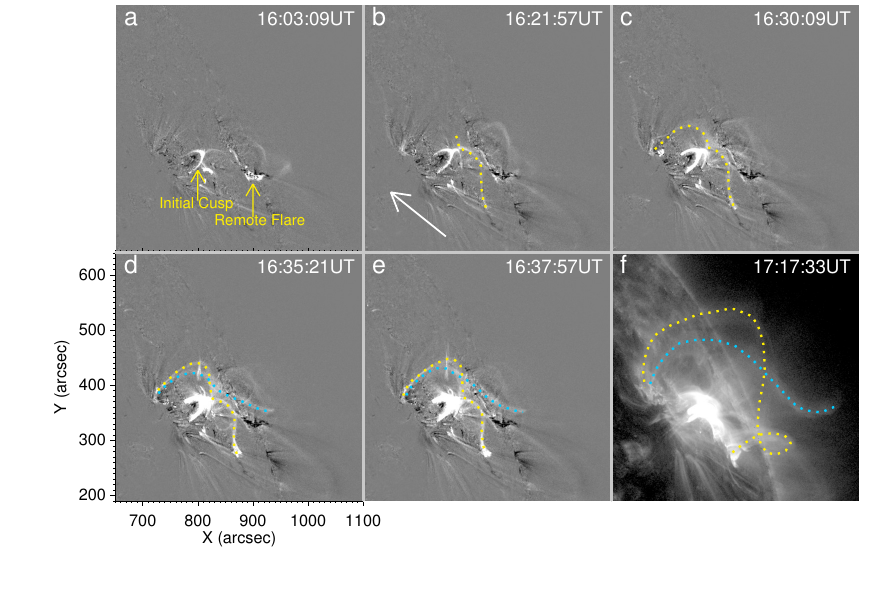}}\vspace*{-0.06\textwidth}
\caption{\textbf{Build up of pre-eruptive MFR.} The AIA 131~{\AA} difference image (subtracting the image taken at 15:50~UT) showing the build up of the pre-eruptive MFR and its relation to the cusp-shaped precursor loops. The two flux bundles of the pre-eruptive MFR are delineated by two curves in yellow and blue (d--f). The oblique arrow in panel b shows the orientation of the main PIL. The animation 1 that starts at 2012 March 13 16:00 UT and ends at 17:59 UT is available online to show the detailed evolution of the slowly rising MFR and precursor loops with the duration of 10 s.} \label{f_formation}
\end{figure*}

The long precursor of interest caused evident coronal emissions at different AIA bands as shown in Figure \ref{f_flare}b. The GOES SXR 1--8 {\AA} flux started to increase at $\sim$16:00~UT and reached a plateau for 20 min. After that, it was further enhanced. At $\sim$16:50~UT, the SXR flux reached a peak and then decreased slightly followed by the onset of the main phase. This evolution at the X-ray band is very similar to that of the integrated intensity of the AIA high-temperature 131 {\AA} passband. In contrast, for the AIA low-temperature 304 {\AA} passband (peaking at roughly 80 kK), only a slow increase in the integrated intensity is observed except for some small fluctuations. The distinction indicates that the energy release process primarily occurred in the corona. Figure \ref{f_flare}b also displays the evolution of the SXR radiation rate and temperature, which are estimated based on fluxes at the two X-ray bands of GOES. One can see that the temperature is mostly above 8 MK during the precursor phase, consistent with the similarity between the evolutions of the SXR 1--8 {\AA} flux and integrated 131 {\AA} intensity. The peak temperature of the precursor appeared at $\sim$16:40~UT, preceding that of the GOES SXR 1--8 {\AA} flux and radiation rate by about 5 min, implying that the plasma is first heated and then induces an enhanced radiation as what happens during the main flare phase \citep{sun14_dem}.

\begin{figure*}
      \centerline{\hspace*{0.00\textwidth}
      \includegraphics[width=0.7\textwidth,clip=]{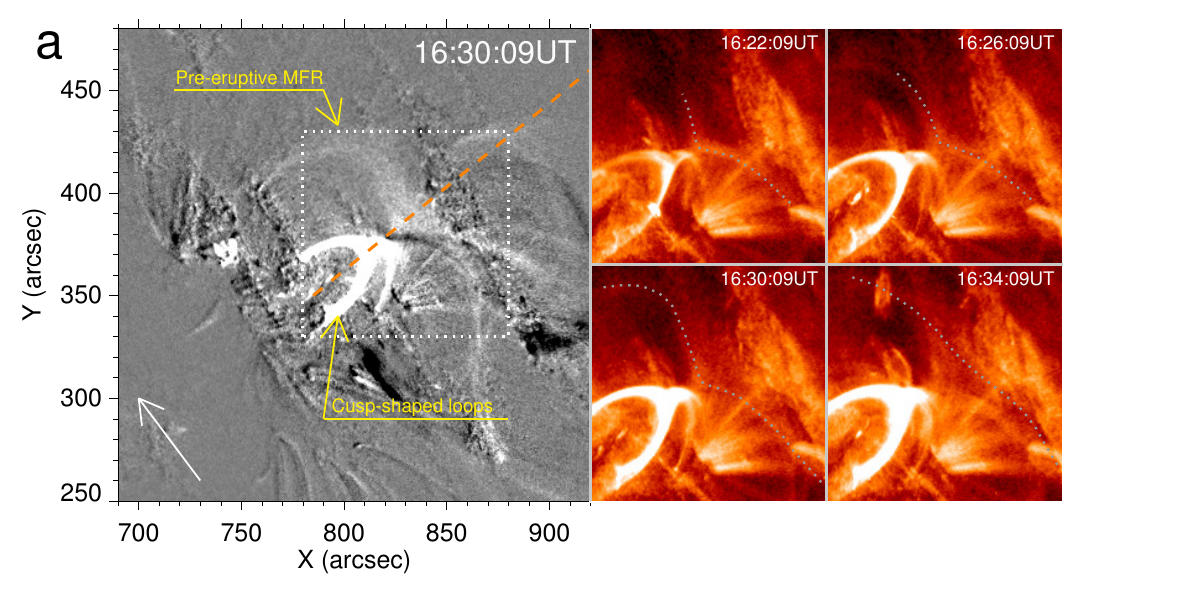}}
      \centerline{\hspace*{0.0\textwidth}
      \includegraphics[width=0.7\textwidth,clip=]{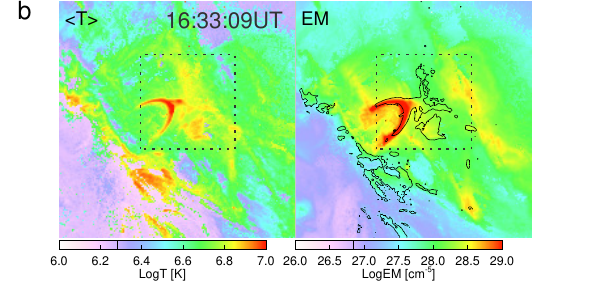}} \vspace*{-0.02\textwidth}
\caption{\textbf{Cusp-shaped precursor loops and its relation to pre-eruptive MFR.} a. The AIA 131~{\AA} difference image (subtracting the image taken at 16:16~UT) showing the cusp-shaped precursor loops and pre-eruptive MFR (left). The rise and change in morphology of the MFR threads are indicated by the curves in zoom-in images (right). The oblique dashed line indicates the main eruption direction. The oblique arrow shows the orientation of the main PIL. b. DEM-weighted average temperature (left) and total EM (right) maps. The temperature contours (black) of 7 MK (corresponding to a logT of 6.85) with an ``X" shape configuration are also overplotted in the EM map. The dashed boxes in black outline the field-of-view of the right of panel a. The animation 2 that starts at 2012 March 13 16:00 UT and ends at 17:00 UT is available online to show the evolution of the MFR threads as shown in the right portion of panel a with the duration of 12 s.} \label{f_recon}
\end{figure*}

Thanks to the capability of the AIA 131 {\AA} and 94~{\AA} passbands to image hot plasma, it is disclosed that the formation of the pre-eruptive MFR involved that of two sets of hot flux bundles. The cusp-shaped hot structure, also named as precursor loops hereafter, is found to be the main source of the precursor emission (Figure \ref{f_formation} and animations 1 and 2). Its activation could be related to a small flare occurred at the nearby neighboring AR 11430 (Figure \ref{f_formation}a). At $\sim$16:00~UT, the first set of relatively diffuse hot threads gradually showed up with their middle being concave and connecting to the top of the cusp-shaped structure. They were almost aligned with the direction of the main PIL and much longer than the precursor loops, presenting an ``M" shape in morphology at $\sim$16:30~UT (Figure \ref{f_formation}c). During the formation, the middle of the hot threads also rose up slowly, being separated from the top of precursor loops, and then became flat (Figure \ref{f_recon}a). At $\sim$16:30~UT, the second flux bundle started to appear with the left footpoints almost being mixed with that of the first one but with the right footpoints far away from the AR 11429. The visibility of the remote dimming at the AR 11430 indicates that the second flux bundle connects the ARs 11429 and 11430 (Figure \ref{f_eruption}c1--c2). The two sets of flux bundles constitute the pre-eruptive channel-like MFR with a bifurcated right leg \citep[also see][]{zhongz19}. Afterwards, the two sets of threads rose up as a whole continuously. This is different from a rise following by a descent of the erupting flux detected in confined eruptions \citep{patsourakos13,liulj18}. The continuous rise of the pre-eruptive MFR also caused a slight amplification of the reconnection as indicated by the brighter precursor loops (Figure \ref{f_formation}) and the increases of the SXR flux, temperature and integrated 131~{\AA} intensity (Figure \ref{f_flare}b). However, comparing with the flare main phase, the SXR flux was still one order of magnitude smaller, indicating that the reconnection still proceeded in a gentle way. Nevertheless, although being moderate, it was critical for the pre-eruptive MFR to be formed and heated (Figure \ref{f_recon}a).

The temperature map shows that the average temperatures of the pre-eruptive MFR and precursor loops are about 8 MK and 10 MK, respectively (Figure \ref{f_recon}b). They are in agreement with the average temperature of the full flaring region estimated from the ratio of two GOES SXR broadband (0.5-4 {\AA}  and 1-8 {\AA}) fluxes \citep[see][]{Thomas85}. Moreover, we find an interesting X-shaped high-temperature plasma structure prior to the eruption, highly resembling to the structure of magnetic reconnection during the eruption \citep{cheng18_cs,chenbin20}. It consists of the upper part of the cusp-shaped loops and the middle of the M-shaped MFR as shown by the contour of 7 MK (right panel of Figure \ref{f_recon}b). Its projected height is $\sim$20 Mm. The two features suggest that magnetic reconnection takes place during the precursor phase and in a high altitude to build up and heat the pre-eruptive structures.

\begin{figure*}
      \centerline{\hspace*{0.00\textwidth}
      \includegraphics[width=0.8\textwidth,clip=]{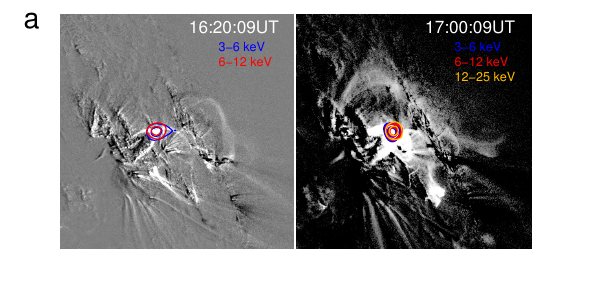}}\vspace*{-0.04\textwidth}
      \centerline{\hspace*{-0.04\textwidth}
      \includegraphics[width=0.75\textwidth,clip=]{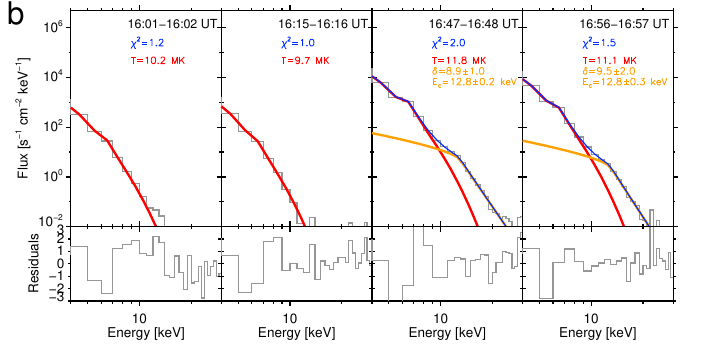}}\vspace*{-0.02\textwidth}
\caption{\textbf{X-ray emission and spectra.} a. RHESSI X-ray source in the energy ranges of 3--6 keV (blue), 6--12 keV (red) and 12--25 keV (yellow) overlaid on the AIA 131~{\AA} difference images (subtracting the image taken at 15:50~UT) showing the energy release locations by precursor reconnection. The two contours for each energy band denote 50\% and 80\% of their maximum emissions. The image on the left and right is scaled linearly and logarithmically, respectively. b. Spectra of X-ray emission and their temporal evolution. The curves in blue display the best fitting to background-subtracted spectra (curves in grey) with the ones in red and yellow indicating the thermal and non-thermal thin-target model, respectively. The resulting fitting residuals are shown in the bottom panels.} \label{f_sxr}
\end{figure*}

\subsection{Locating X-ray emissions}

The RHESSI hard X-ray (HXR) data show that, during the time period of 16:00-16:20~UT, the HXR emissions only appeared in the energy bands below 12 keV and were from the source concentrated at the top of precursor loops, as shown in Figure \ref{f_sxr}a. The left two panels of Figure \ref{f_sxr}b display that the corresponding X-ray spectra at $\sim$16:01 and 16:15 UT can be well fitted by a thermal model. It gives a thermal temperature of $\sim$10 MK, very similar to the DEM-average temperature of the region at the top of precursor loops as derived from the DEM analysis, indicating pertinent reconnection process being of thermal-dominated. As the pre-eruptive MFR formed, the HXR emissions at the top of the precursor loops were gradually enhanced. At $\sim$16:47 UT, the emissions in the energy range of 3--6 keV and 6--12 keV increased by almost one order of magnitude relative to the value observed half an hour earlier (Figure \ref{f_sxr}b). At the same location, the emission in the higher energy range (e.g., 12--25 keV) also appeared (the right two panels of Figure \ref{f_sxr}a), suggestive of a non-thermal property. The combination of a thermal model and a thin-target model best fits the X-ray spectra at two following times (middle and right panels of Figure \ref{f_sxr}b). The fitting gives a thermal temperature of 11-12 MK and a cut-off energy of 12.8 keV for the non-thermal electrons. This indicates that magnetic reconnection in the later stage of the precursor phase was also capable of accelerating electrons.
However, the accelerated electrons still have a relatively low energy, mostly $<$20 keV, showing that the reconnection process was not energetic, probably similar to that during micro-flares as recently observed by the Spectrometer/Telescope for Imaging X-rays (STIX) onboard Solar Orbiter \citep{Battaglia21}.

\subsection{Initiation of MFR-induced CME}

\begin{figure*}
      \vspace*{-0.10\textwidth}
      \centerline{\hspace*{0.00\textwidth}
      \includegraphics[width=0.8\textwidth,clip=]{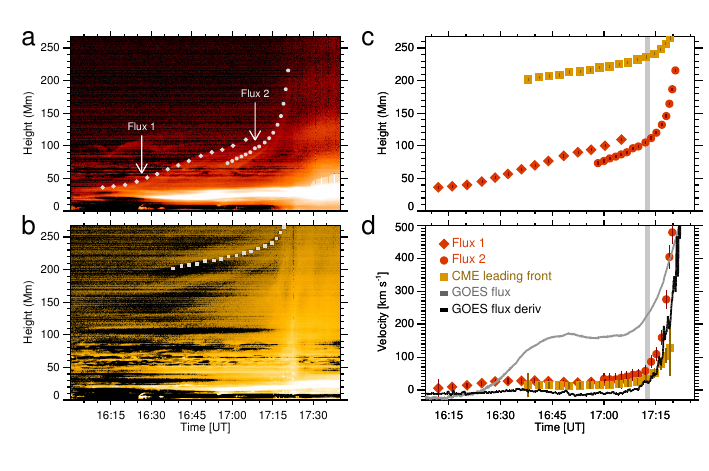}}\vspace*{-0.04\textwidth}
      \centerline{\hspace*{0.0\textwidth}
      \includegraphics[width=0.7\textwidth,clip=]{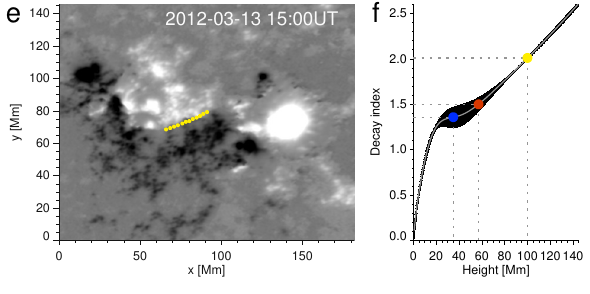}}\vspace*{-0.02\textwidth}
\caption{\textbf{Early kinematics and initiation of pre-eruptive MFR.} a. Slice-time plot of the AIA 131~{\AA} difference images showing the evolution of two flux bundles of the pre-eruptive MFR. Diamonds and filled circles indicate their height-time measurements. b. Same as panel a but for the AIA 171~{\AA} passband representing the expanding overlying field. c. Temporal evolution of the heights of the MFR (red) and expanding overlying field (yellow). d. Temporal evolution of the velocities overplotted by the GOES SXR 1--8~{\AA} flux (gray) and its time derivation (black). The uncertainty in velocity is mainly from that in height, which is estimated to be 2 Mm. The vertical slits in panel c and d indicate the onset of the CME impulsive acceleration and flare main phase with their width denoting the uncertainty (one minute). e. The radial component of HMI Cylindrical Equal-Area (CEA) vector magnetogram. The white (black) indicates the magnetic field upward (downward). f. The decay index distribution with height above the PIL, which is an average of all height-profiles of the decay index at the yellow dots as shown in panel e. The bars in grey denote its uncertainty as derived by the standard deviation of height-profiles. The dots in blue and yellow represent the initial and onset height (35 and 100 Mm) of the MFR axis, respectively. Their reference point is the midpoint of the line segment connecting the two footpoints of the cusp-shaped precursor loops. The dot in red points out the the critical decay index of 1.5 for a ring current.} \label{f_initiation}
\end{figure*}

The temporal variations of the height and velocity of the pre-eruptive MFR provide valuable information to disentangle initiation models. Figure \ref{f_initiation}a shows the long-duration slow rise of the pre-eruptive MFR consisting of two rising flux bundles. After $\sim$17:00 UT, the MFR gradually became vague because of its expansion. During the entire precursor phase, as a response to the rise of the hot MFR, the overlying field also gradually expanded, as indicated by diamonds in Figure \ref{f_initiation}b, and then evolved toward the CME leading front during the main phase.  

Figure \ref{f_initiation}c--d display that the MFR velocity increased very slowly in the precursor phase, varying from $\sim$5 km s$^{-1}$ at $\sim$16:15 UT to $\sim$25 km s$^{-1}$ at $\sim$16:40 UT. The average acceleration was only $\sim$13 m s$^{-2}$. Afterwards, the MFR even started to slow down, the velocity decreased from $\sim$25 km s$^{-1}$ at $\sim$16:40 UT to $\sim$20 km s$^{-1}$ at $\sim$16:55 UT, with a deceleration of about --6 m s$^{-2}$. At $\sim$17:00 UT, because of the second flux bundle, the MFR velocity increased again. The temporal evolution of the velocity of the pre-eruptive MFR during the whole precursor phase roughly kept in step with the variation of the GOES 1--8 {\AA} SXR flux. For the CME leading front, its velocity kept a small value ($\sim$14 km s$^{-1}$) before eruption at $\sim$17:12 UT. However, as the main phase began, the velocity of the CME leading front increased to $\sim$110 km s$^{-1}$ in 6 minutes. The MFR was accelerated more impulsively, the velocity increased from $\sim$50 km s$^{-1}$ at $\sim$17:12 UT to $\sim$450 km s$^{-1}$ at $\sim$17:20 UT with an acceleration of $\sim$830 m s$^{-2}$, almost two orders of magnitude larger than that during the precursor phase. Meanwhile, the SXR flux also increased impulsively (see its time derivative), in synchronisation with the variation of the MFR velocity.
 
The MFR fast acceleration may be triggered by the torus instability, which occurs when the decay of the background magnetic field with height exceeds a threshold \citep{kliem06,fan07,aulanier10}. From the distribution of the decay index as shown in Figure \ref{f_initiation}f, one can see that the decay index first quickly and then gradually increases with height. At the onset time of the impulsive acceleration (17:13 UT), the upper and lower edges of the MFR reached heights of $\sim$110 and $\sim$90 Mm, respectively. The average ($\sim$100 Mm) of them is regarded as the height of the MFR axis, where the decay index is found to be $\sim$2.0, obviously exceeding all critical values of torus instability derived theoretically \citep{kliem06,demoulin10}. This shows that the torus instability has occurred, probably earlier than 17:13 UT, because it takes time to accumulate speed in an exponentially-accelerating instability starting from a weak perturbation to overtake the slow-rise velocity caused by the independent precursor reconnection. 

We also inspect the decay index during the slow rise phase. For the acceleration stage of the precursor phase (16:15--16:40 UT), the height of the MFR axis was below 60 Mm, the corresponding decay index was mostly smaller than $\sim$1.5, the critical value for a toroidal current ring \citep{kliem06}, and the statistical average of critical decay indices for AR eruptions \citep{cheng20}. Furthermore, the deceleration of the MFR during the following 15 minutes conflicts with an expected exponential acceleration during the early development stage of torus instability \citep{torok05,schrijver08_filament}. Thus, although the decay index keeps increasing as the MFR is elevated continuously, the torus instability seems to not take effect in the most of precursor phase. In contrast, once entering the main phase, the temporal variation of the MFR height is found to exactly follow an exponential form (Figure \ref{f_fit}). These results support that the torus instability plays a critical role in initiating the MFR impulsive acceleration and fast flare energy release, or in other words, it is the key to turn the moderate reconnection in the precursor phase to the runaway reconnection in the main phase.

\subsection{Driver of MFR Slow-rise}

To understand the dominant driving mechanisms underneath the long-term slow rise of the MFR and the transition toward the eruption, as well as the relations of various involved features to the inferred moderate precursor reconnection, we run a zero-$\beta$ three-dimensional (3D) magnetohydrodynamic (MHD) simulation. To broadly compare with observations, in our numerical model, the initial magnetic configuration consists of an asymmetric bipolar field. It is driven by means of line-tied shearing motions at the bottom boundary, which are often observed near the PIL. The flux cancellation is also introduced by magnetic diffusion at the bottom boundary. The parameter setups are the same as \citet{aulanier12} only with a higher spatial resolution (375$\times$375$\times$336). 
\begin{figure*}
      \vspace*{-0.0\textwidth}
      \centerline{\hspace*{0.00\textwidth}\vspace*{0.02\textwidth}
      \includegraphics[width=0.8\textwidth,clip=]{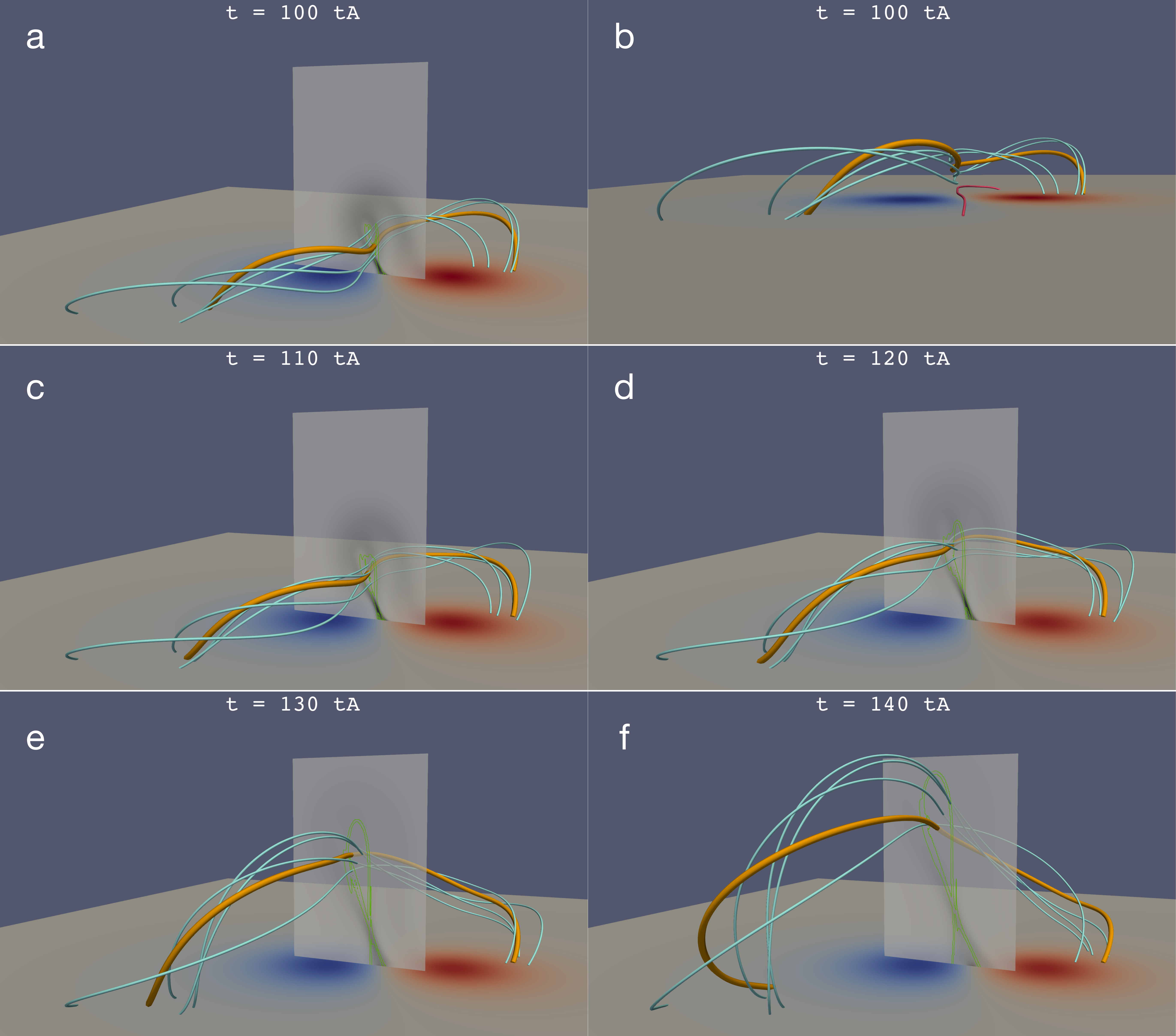}}
\caption{\textbf{3D magnetic field lines at five different times displaying the slow rise and early eruption of the MFR}. The blue and orange tubes show the MFR field lines, the red tube in panel b shows one precursor loop. The bottom images show the vertical magnetic field components at the bottom boundaries of the MHD simulation domain. The vertical planes (y=-0.06) perpendicular to the MFR axis show the distribution of current density $j$ with the contours in green indicating $\log Q=3$. The onset time of the MFR eruption is in period of 120-125$t_A$. The animation 3 is available online to show the evolution of 3D M-shaped field lines during the early rise phase with the duration of 46 s.} \label{f_fieldline}
\end{figure*}

The 3D MHD simulation shows that, as the bald-patches (BPs) bifurcates into a quasi-separatrix layer (QSL) containing a coronal hyperbolic flux tube \citep[HFT;][]{titov02} lying below the MFR, the MFR is slowly evolving before it reaches an eruptive stage. The HFT reconnection is believed to inject newly formed flux into the MFR, making it ascend further \citep{aulanier10}. Although the MFR in the simulation only includes one set of helical threads, many characteristics are still similar to observations, including (1) that the newly reconnected flux first presents an ``M" shape and then becomes flat (orange field lines in Figure \ref{f_fieldline}a-\ref{f_fieldline}d); (2) that the precursor reconnection forming the ``M"-shaped MFR field lines and precursor loops is not energetic prior to the main eruption; (3) that the MFR eruption does not start until its axis reaches an altitude where the decay index of the background field is large enough to allow the occurrence of ideal torus instability (Section \ref{mhd}); and (4) that the height of the erupting MFR during the early acceleration phase increases exponentially (see Figure \ref{height}).

\begin{figure*}
      \vspace*{0.0\textwidth}
      \centerline{\hspace*{0.0\textwidth}\vspace*{-0.0\textwidth}
      \includegraphics[width=1.\textwidth,clip=]{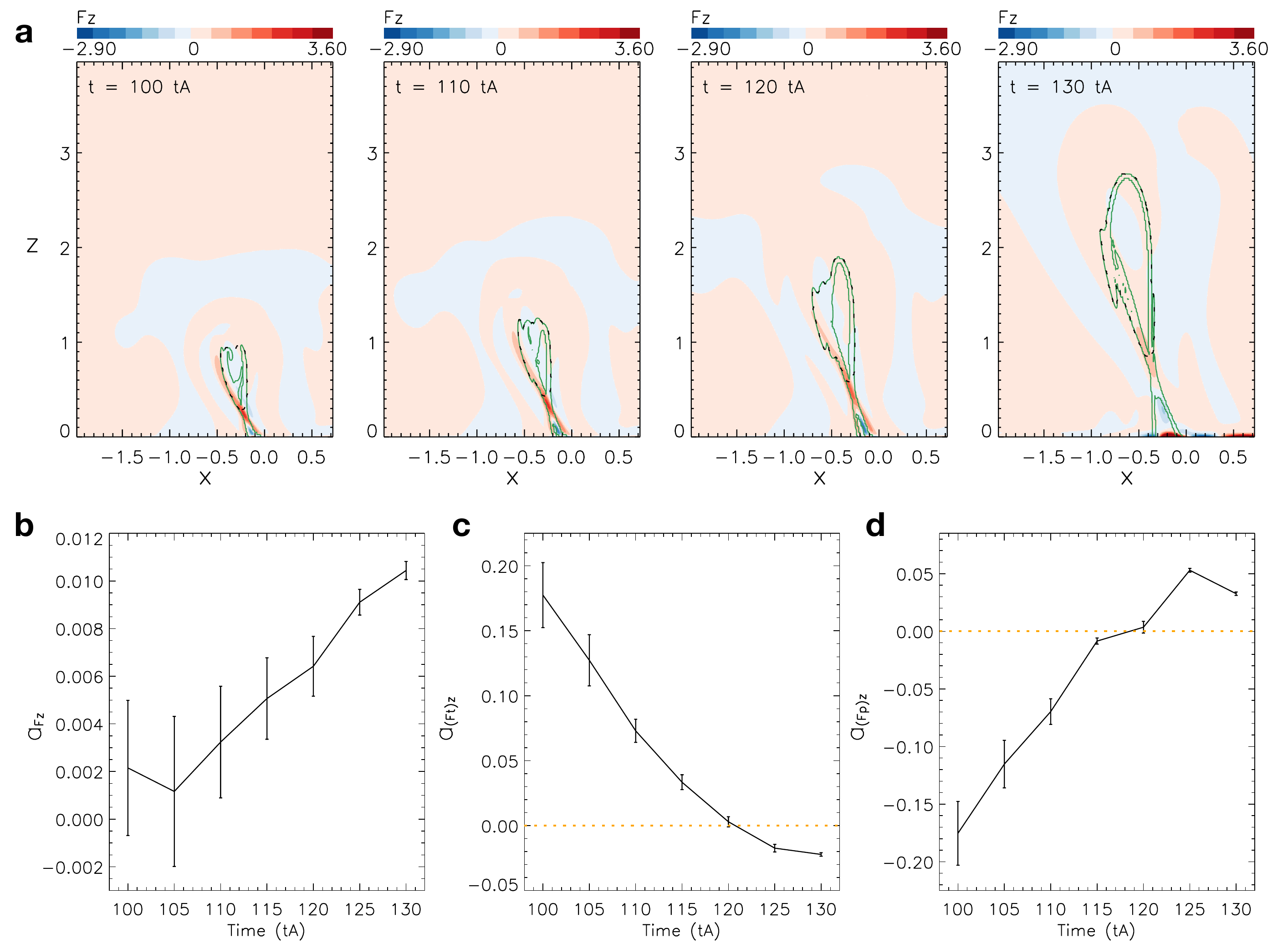}}
\caption{\textbf{Lorentz force during the slow rise and early eruption of the MFR.} a. Distributions of the vertical component of the Lorentz force density $F_z$ at the plane y=-0.06, as shown in Figure \ref{f_fieldline}, at four different times. The boundaries of the MFR are delineated by the dashed lines; the contours of $\log Q=3$ are shown by the curves in green. b--d. Temporal evolution of the MFR accelerations ($a_{F_z}$, $a_{(F_t)_z}$, $a_{(F_p)_z}$) contributed by the vertical components of the Lorentz force $F_z$, magnetic tension $(F_t)_z$ and magnetic pressure gradient $(F_p)_z$, respectively. Their errors indicated by vertical bars are mainly from the uncertainty in determining the MFR outer boundary, which is achieved through changing $\log Q$ value from 2.4 to 3.6. The horizontal dotted lines indicate zero acceleration.} \label{f_force}
\end{figure*}

The MHD simulation enables disclosing the driving forces acting in the slow rise of the pre-eruptive MFR and the following acceleration phase, respectively. The distribution of the vertical component of the Lorentz force in the plane perpendicular to the MFR axis and crossing the HFT show that the Lorentz force near the HFT and the outer part of the MFR is pointing upward but that in the central part of the MFR is mostly directed downward (Figure \ref{f_force}a). Nevertheless, the downward force within the MFR is negligible, so that the net force of the MFR is obviously dominated by the upward one which drives the MFR to rise up slowly with a small acceleration (Figure \ref{f_force}b). On the other hand, the Lorentz force below the HFT is mainly pointing downward, which causes shrinkage of precursor loops. The Lorentz force is further decomposed into two components (magnetic tension and pressure gradient). One can find that, during the slow rise phase of the MFR, the upward directed Lorentz force producing a positive acceleration is primarily contributed by the upward directed magnetic tension (Figure \ref{f_force}c), in agreement with the observation that the middle part of the ``M"-shaped flux tends to rise up obviously and gradually becomes flat. As the fast eruption starts, the acceleration induced by the upward Lorentz force quickly increases, explaining the fast acceleration of the MFR eruption as observed. However, the driving force for the large acceleration is no longer the magnetic tension but the magnetic pressure gradient. The latter gradually changes from negative to positive and significantly counteracts the negative magnetic tension as the eruption enters into the main phase (Figure \ref{f_force}d). 

\section{Summary and discussions} 
In this paper, we present comprehensive observations of a long-lasting precursor phase before a major solar eruption. With further combination of suitable viewing angle and multi-wavelength data, it is disclosed that the heating and slow rise of the pre-eruptive hot MFR are achieved through the precursor reconnection as often speculated previously \citep{wanghaimin17,zhou_17,Awasthi18,chenhc_2019,HernandezPerez19,gou19}. The precursor reconnection is found to take place at the X-shaped high-temperature plasma sheet. The continual formation of the ``M"-shaped hot threads via the precursor reconnection results in the heating and early rise of the entire MFR, as well as the formation of precursor loops. It is surprising that the main features generally observed in the impulsive phase have counterparts in the precursor phase that are physically linked to the precursor reconnection. However, both the slow rise of the pre-eruptive hot MFR ($<$30 km s$^{-1}$) and the relatively low energy of accelerated electrons ($<$20 keV) suggest that the precursor reconnection is far less efficient than that in the main eruption phase \citep[e.g.,][]{cheng18_cs}.

In spite of being moderate, the precursor reconnection is critical for lifting the MFR along an equilibrium sequence in a mutual feedback process. On the one hand, the precursor reconnection forms the M-shaped flux that accumulates and rises up as illustrated in Figure \ref{bottom_boundary}. On the other hand, the rising MFR slightly enhances the reconnection that can further inject flux to the MFR. Such a feedback is strongly indicated by the simultaneity between the increases in the height of the pre-eruptive MFR and the enhancement of associated SXR emissions. The transition from moderate reconnection to fast reconnection is switched on by the fast acceleration of the MFR. This is most likely caused by the ideal torus instability as the transition from the slow rise to the fast acceleration of the MFR occurs at the height where the decay index of the background field exceeds the thresholds of torus instability and that the temporal evolution of the MFR acceleration highly resembles an exponential profile. Once entering the main eruption phase, the moderate reconnection immediately transitions to runaway reconnection responsible for the flare impulsive phase. The fast reconnection rapidly accelerates the CME eruption, which vice versa drives the opposite-directed magnetic fields to continuously participate in the reconnection. With the feedback more efficiently operating in this period, the eruption is finally accelerated to a high speed in about one hour. Moreover, as shown in the MHD simulation, the precursor reconnection occurs between highly sheared arcades \citep[also see][]{aulanier12}, the guide field of the reconnection is thus large, which could be the reason why the precursor reconnection is not that energetic \citep[e.g.,][]{Leake20,Dahlin_2022}. 

The synergism of the moderate-reconnection-formed MFR and ideal torus instability causing the transition from the precursor phase to the main eruption phase as revealed and further testified by our 3D MHD simulation can be used to clarify the applicability of the initiation models proposed in the past decades. The tether-cutting-like topology of the precursor reconnection seems to support the tether-cutting initiation model \citep{moore01}, which, however, is insufficient if working alone without the presence of the torus instability. The tether-cutting reconnection is found to be only efficient once initiated by tearing mode instability as proved by recent numerical simulations \citep{jiang21}. This is at variance with the results of observation-constrained simulation \citep{inoue18} and observational characteristics for the 2012 March 13 event, where the reconnection is found to be moderate before the onset of the fast eruption. 
Moreover, in the event under study, no prominent signatures for the MFR writhing motion are observed during the slow rise phase. Thus, we tend to exclude the possibility of the kink instability \citep{torok03} causing both the precursor phase and the main eruption phase. 
It is worthy of noticing that the first bundle of the pre-eruptive MFR seems to present an untwisting motion in the early phase of the eruption (Figure \ref{f_formation}f). However, after a careful inspection, it is more likely to be the apparent manifestation of the MFR morphology varying from the ``M" to semicircle shape as previously detected \citep{zhang12,cheng13_driver}. In addition, it was suggested that the onset of the eruption may correspond the transition of magnetic field topology that embeds the pre-eruptive MFR, so to say, from BPS to HFT \citep{savcheva12b}. However, in our observations, only an HFT-like configuration appears in the precursor phase. Therefore, as proved in our numerical model, such a topological transition may occur much earlier than the initiation of the eruption, if it exists at all. 

The long-duration precursor is equivalent to a confined flare preceding the following eruptive one, during which both magnetic helicity and twist are thought to be quickly injected to the pre-eruptive MFR \citep{priest17}. However, it presents two characteristics obviously different from those during confined eruptions. First of all, a continuously rising pre-eruptive MFR toward the eruption differs from the MFR during confined flares that initially rises but finally stops at the high corona \citep{liulijuan18,kliem21}. Secondly, the high-temperature of the pre-eruptive MFR in combination with its morphology evolution and the appearance of thermal X-ray source above the precursor loops provide solid evidence for slow reconnection heating during the slow rise prior to the main eruption; while such a process may be unnecessary, even absent \citep{patsourakos13,chintzoglou15}, in the interval between the preceding confined flares and the following successful eruptions.

Lastly, as only one particular event is investigated here, more similar observations and in-depth MHD modelling are suggested in the future to justify the universality of the mechanisms we have determined for the heating and slow rise precursor of pre-eruptive configurations of solar eruptions.

\begin{acknowledgments}
We appreciate all referees who reviewed the manuscript and provided their comments and constructive suggestions. 
We also thank Chun Xia, Bernard Kliem, Jie Zhang, Jun Chen and Lakshmi Pradeep Chitta for their helpful discussions. AIA data are courtesy of NASA/SDO, a mission of NASA's Living With a Star Program. \textsl{STEREO}/SECCHI data are provided by a consortium of NRL (US), LMSAL (US), NASA/GSFC (US), RAL (UK), UBHAM (UK), MPS (Germany), CSL (Belgium), IOTA (France), and IAS (France). X.C., C.X. and M.D.D. are supported by National Key R\&D Program of China under grants 2021YFA1600504 and by NSFC under grant 12127901. X.C. is also supported by Alexander von Humboldt foundation. G.A. and C.X. acknowledge financial support from the French national space agency (CNES), as well as from the Programme National Soleil Terre (PNST) of the CNRS/INSU also co-funded by CNES and CEA.
\end{acknowledgments}

\appendix
\section{DEM inversion}
The differential emission measure (DEM) is reconstructed by the ``xrt\b{ }dem\b{ }iterative2.pro" routine in the Solar Software (SSW) using six co-aligned AIA EUV images. The observed intensity $I_{i}$ for the passband ${i}$ can be written as:
 \begin{equation}
 {I_{i}}  =  \int DEM(T) \times R_i (T) \mathrm{d}T + \delta I_{i},
 \end{equation}
where $DEM(T)$ denotes the plasma DEM, $R_{i}(T)$ is the temperature response function and $\delta I_{i}$ is the uncertainty in intensity $I_{i}$. The temperature range of inversion is set as 5.5$\leq$ log${T}\leq$ 7.5. 

We calculate the average temperature and total EM by means of the following two formulae:
\begin{equation}
\label{T}
\bar{T}= \frac{\int DEM(T) \times T dT}{\int DEM(T) dT} \\
\end{equation}

\begin{equation}
\label{EM}
EM= \int DEM(T) dT.\\
\end{equation}
We also ran 100 Monte Carlo (MC) simulations by adding a random noise corresponding to the uncertainties of the observed intensities, derived by ``aia\b{ }bp\b{ }estimate\b{ }error.pro" in SSW, to the intensity $I_{i}$ and then resolving the DEM. It is found that in the temperature range of 5.7$\leq$ log${T}\leq$ 7.4, which is selected to integrate Equation (\ref{T}) and (\ref{EM}), the 100 MC solutions are well constrained.

\section{Dimmings}
\begin{figure*} 
      \vspace{-0.0\textwidth}
      \centerline{\hspace*{-0.04\textwidth}
      \includegraphics[width=0.7\textwidth,clip=]{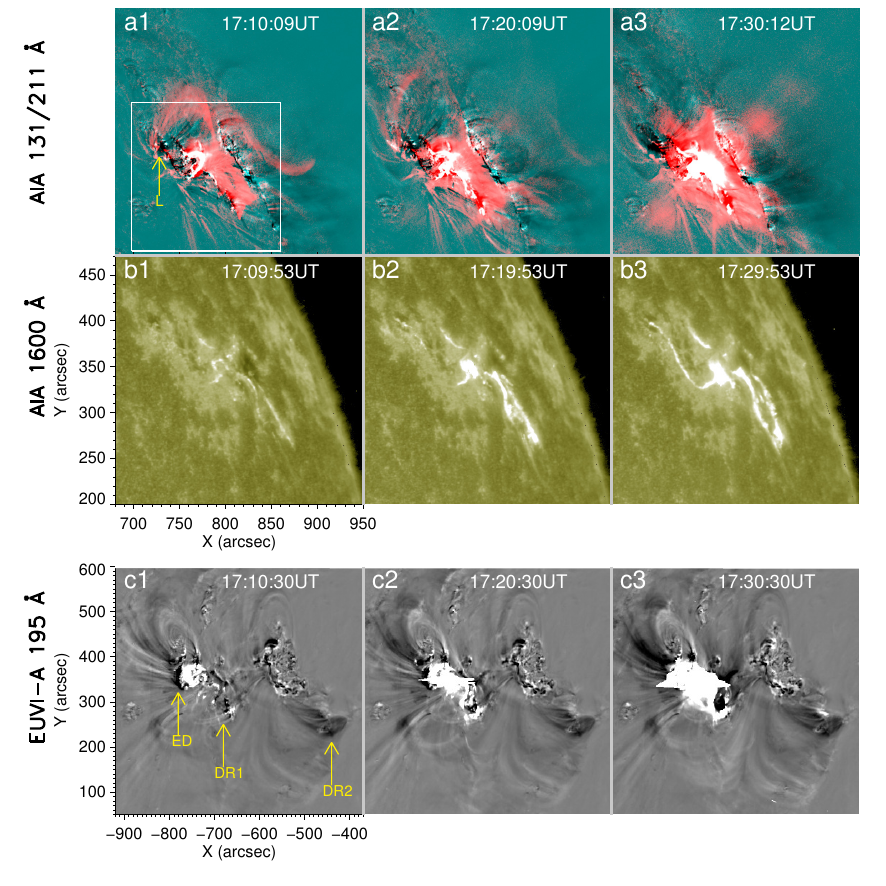}}\vspace*{-0.03\textwidth}
\caption{a1--a3. Time-sequence of composite of the AIA 131~{\AA} (red) and 211~{\AA} (cyan) difference images showing the early rise and eruption of the pre-eruptive MFR and induced flare. b1--b3. The AIA 1600~{\AA} images displaying the flare ribbons, the field-of-view is indicated by the box in panel a1. c1--c3. The EUVI 195~{\AA} difference images (subtracting the image at 14:00 UT) representing the dimming regions, which include the footpoints of the erupting MFR. The left footpoint (L) is indicated in panel a1, the left dimming (ED) and right two dimmings (DR1 and DR2) are pointed out by the arrows in panel c1.} \label{f_eruption}
\end{figure*}

Figure \ref{f_eruption} shows the evolution of the three dimming regions and flare ribbons from the two perspectives of SDO and STEREO-A before and during the eruption. It is found that the east dimming has appeared prior to the main eruption, as seen in the SDO-AIA difference images (Figure \ref{f_eruption}a1). Not surprisingly, the left footpoints of the pre-eruptive MFR were cospatial with the east dimming (ED in Figure \ref{f_eruption}c1). Considering that the pre-eruptive MFR consists of two sets of threads that have a different connectivity on the right, their footpoints are expected to correspond to two dimmings, which is confirmed by the EUVI-A 195 {\AA} running-difference images. It is revealed that the one is on the right of the main flare loops and the other is near the AR 11430. As the MFR took off, the two dimming regions expanded outward and further darkened, implying a plasma rarefaction caused by the eruption (Figure \ref{f_eruption}a2-\ref{f_eruption}a4). The erupted fluxes should be mostly from NOAA 11429 as where the main dimmings were observed (ED and DR1 in Figure \ref{f_eruption}c1).

\restartappendixnumbering

\section{Kinematical analyses}
To measure the MFR height, we take a slice along the MFR eruption direction (Figure \ref{f_recon}a) and make slice-time plots of the AIA 131 {\AA} and 171 {\AA} base-difference images (Figure \ref{f_initiation}a--\ref{f_initiation}b). Based on the slice-time plots, we measure projected heights of the MFR upper edge and CME leading front as shown in Figure \ref{f_initiation}a--\ref{f_initiation}c. The projected heights are simply corrected assumes that the eruption is along the radial direction during the early phase (as indicated by the oblique dashed line in Figure \ref{f_recon}a). With the first order numerical derivative, we then calculated the velocities of the MFR and CME leading front. 

\begin{figure*}
      \centerline{\hspace*{0.00\textwidth}
      \includegraphics[width=0.9\textwidth,clip=]{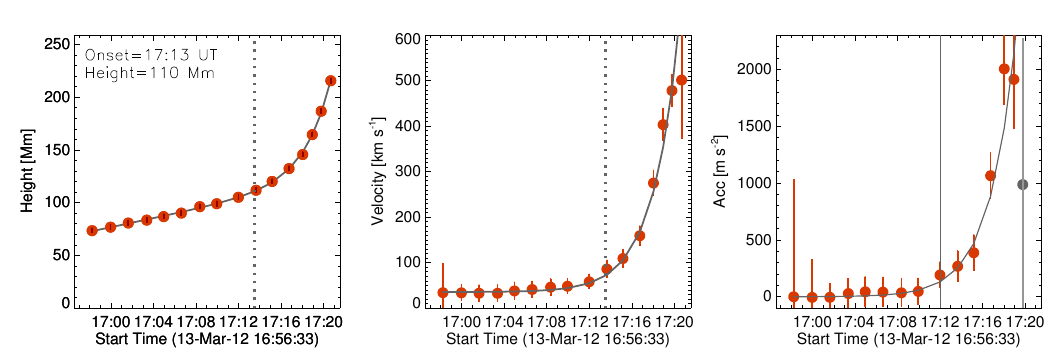}}\vspace*{-0.0\textwidth}
\caption{Model fitting for temporal evolution of the height (left), velocity (middle) and acceleration (right) of the second MFR bundle. The dots in red show measured data points with the vertical bars indicating the uncertainties. The curves in grey show the fitted results. The vertical dotted lines in the left and middle panels represent the onset time of the main acceleration phase derived by the model. The solid line in the right panel shows the onset directly from the acceleration-time profile.} \label{f_fit}
\end{figure*}

By comparing multiple fit functions, it was found that, for the majority of events, the height-time profiles of solar eruptions in the lower corona can be best fitted by the function:
\begin{equation}
\label{fun}
h(t) = a \exp(bt) + ct + d, \\
\end{equation}
which is a superposition of a linear and exponential component, mimicking the slow-rise phase and the early impulsive acceleration phase, respectively \citep{cheng20}. Here, we take advantage of the superposed function to fit the height-time profile of the second MFR bundle that continuously evolved from the precursor to the main phase. Figure \ref{f_fit} shows that the measured height-time data are perfectly fitted by Equation \ref{fun}. Furthermore, all velocities, even accelerations, that are directly calculated by numerical derivative of height-time data are also found to follow the derivative curve of the fit function with a very small discrepancy except for the last point. 

In terms of the best fitted function, we further estimate the onset of the impulsive acceleration phase, i.e., the break-point time where the velocity of the exponential term starts to dominate (equal) that of the linear term. This gives an onset time of $\sim$17:13 UT. Moreover, we also estimate directly from the acceleration-time profile the onset time, i.e., when the acceleration begins to obviously increase at $\sim$17:12 UT. The onset times we derived by two different methods are synchronized with that of the main flare phase (17:12 UT). For more details on determination of the eruption onset and its uncertainty, the reader can refer to \citet{cheng20}. 

\restartappendixnumbering

\section{3D coronal magnetic extrapolation}
Based on a potential field model, we extrapolate the 3D coronal magnetic field by the Green function method using the radial component of the HMI vector field as shown in Figure \ref{f_initiation}e as the bottom boundary. The constrained background field over the erupting MFR is approximated by the horizontal component of extrapolated potential field. The decay index of the background field is calculated as follows
\begin{equation}
\centering
n(h)=-\frac{d(\ln B_\mathrm{t})}{d(\ln h)},
\end{equation}
where $B_\mathrm{t}$ and $h$ denote the horizontal component of the background field and the height above the photosphere, respectively. 

The accuracy of the extrapolated 3D coronal potential field largely depends on that of the bottom boundary, which is selected as the radial component of the HMI vector magnetogram at 15:00 UT on 2012 March 13. The 180$^{\circ}$ ambiguity in the horizontal component is removed using a minimum energy method. In addition, the data are reprojected from helio-projective Cartesian to heliographic Cylindrical Equal-Area (CEA) coordinates. Finally, it is worth mentioning that the measured vector field close to the solar limb is still less accurate than that near the disk center. This may influence the accuracy of the calculated background field and thus the decay index. However, as discussed previously \citep{cheng20}, the decay property of the background field is primarily determined by the large-scale structure of ARs, which generally evolve slowly after their emergence. The influence is thus not expected to be fatal.

\section{The MFR Geometry}
The thresholds of an MFR taking place torus instability are closely related to its geometry. For a straight and toroidal thin current ring, two particular limited cases, the critical decay index of the background field was deduced to be 1 and 1.5, respectively \citep{kliem06,demoulin10}. For the current 2012 March 13 event, the pre-eruptive MFR presents a curved loop-like structure, deviating from a full torus. Moreover, the pre-eruptive MFR is composed of two flux bundles with the right footpoints being located at the different regions. This is essentially a result of the nonuniform distribution of the MFR current. The two factors may influence the critical value of torus instability, however, based on \citet{kliem06}, which is not expected to be significant.

\section{The MHD model}\label{mhd}
We run a zero-$\beta$ MHD simulation performed by the Observationally driven High-order Magnetohydrodynamics code \citep[OHM;][]{aulanier05,aulanier10}. The simulation starts from an asymmetric bipolar potential field. To drive the potential field evolving toward a highly sheared state, we impose the driving motion at the bottom boundary, which is mainly manifested as two shearing flows on two sides of the main PIL, reaching its maximum close to the PIL and has little effect in the center of each polarity. In addition, since the driving motion follows the contours of $B_z$, the vertical component of the magnetic field at the bottom boundary is hardly changed by this motion.

The simulation is composed of the shearing phase with shearing motions imposed and the relaxation phase without driving motions at the bottom boundary. During the shearing phase, the flux cancellation is achieved by adding a photospheric resistivity $\eta^{phot}=1.44\times10^{-3}$ on the bottom; an uniform coronal resistivity, $\eta=4.8\times10^{-4}$, is set in the whole domain except the bottom boundary. During the relaxation phase, the photospheric resistivity is set to be zero, and the coronal resistivity is multiplied by 4 during the eruption for numerical stability.

\begin{figure*}
      \vspace*{-1.2\textwidth}
      \centerline{\hspace*{0.4\textwidth}
      \includegraphics[width=1.2\textwidth,clip=]{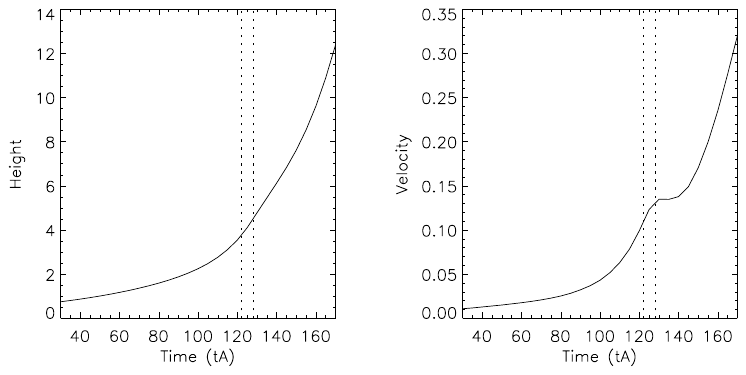}}\vspace{-0.08\textwidth}
\caption{\textbf{Temporal evolution of the height and velocity of the MFR in the simulation.} The height of the MFR is estimated by tracking an overlying field line that is located at the top of the MFR in the direction of the eruption. The vertical dashed lines mark the time interval in which the driving motion at the bottom boundary gradually decreases to zero by following a ramp function.} \label{height}
\end{figure*}

We determine the onset time of the MFR eruption with a series of tests, in which we stop the driving motion at different times and then relax the system. We find that the MFR fails to erupt in a control simulation where the driving motion is switched off at $t=120t_A$ (with an interval of $2\Delta t=6t_A$) but erupts successfully in the simulation where the driving motion is stopped at $125t_A$ as analysed here. Thus, the onset time of the eruption should be in the time period of $120-125t_A$. The decay index at the height of the MFR axis at $t=125t_A$ is found to be close to the theoretical threshold of the torus instability \citep[$\sim$1.5;][]{kliem06}. Afterwards, both the height and velocity of the MFR increase exponentially (Figure \ref{height}). 

\restartappendixnumbering

\section {Determining the MFR boundary}
We investigate the mechanism that drives the slow rise of the MFR by analyzing the integrated z-components of the Lorentz force and its two components (magnetic tension and pressure gradient) at the section of the MFR perpendicular to its axis. Note that, the magnetic tension force (pressure gradient) analyzed here refers to the component of magnetic tension force (pressure gradient) in the normal direction of magnetic field as the tangential component has no contribution to the acceleration.

To identify the MFR boundary, we calculate the squashing degree Q, which measures the mapping of the field lines. The squashing degree Q is defined by \citet{titov02} as:
\begin{equation}
\centering
Q=\frac{(\frac{\partial X}{\partial x})^2+(\frac{\partial X}{\partial y})^2+(\frac{\partial Y}{\partial x})^2+(\frac{\partial Y}{\partial y})^2}{|\frac{\partial X}{\partial x}\frac{\partial Y}{\partial y}-\frac{\partial X}{\partial y}\frac{\partial Y}{\partial x}|},
\end{equation}
where $(x,y)$ and $(X,Y)$ are coordinates of two footpoints of a field line.

\begin{figure*}
      \centerline{\hspace*{0.0\textwidth}\vspace*{-0.0\textwidth}
      \includegraphics[width=0.8\textwidth,clip=]{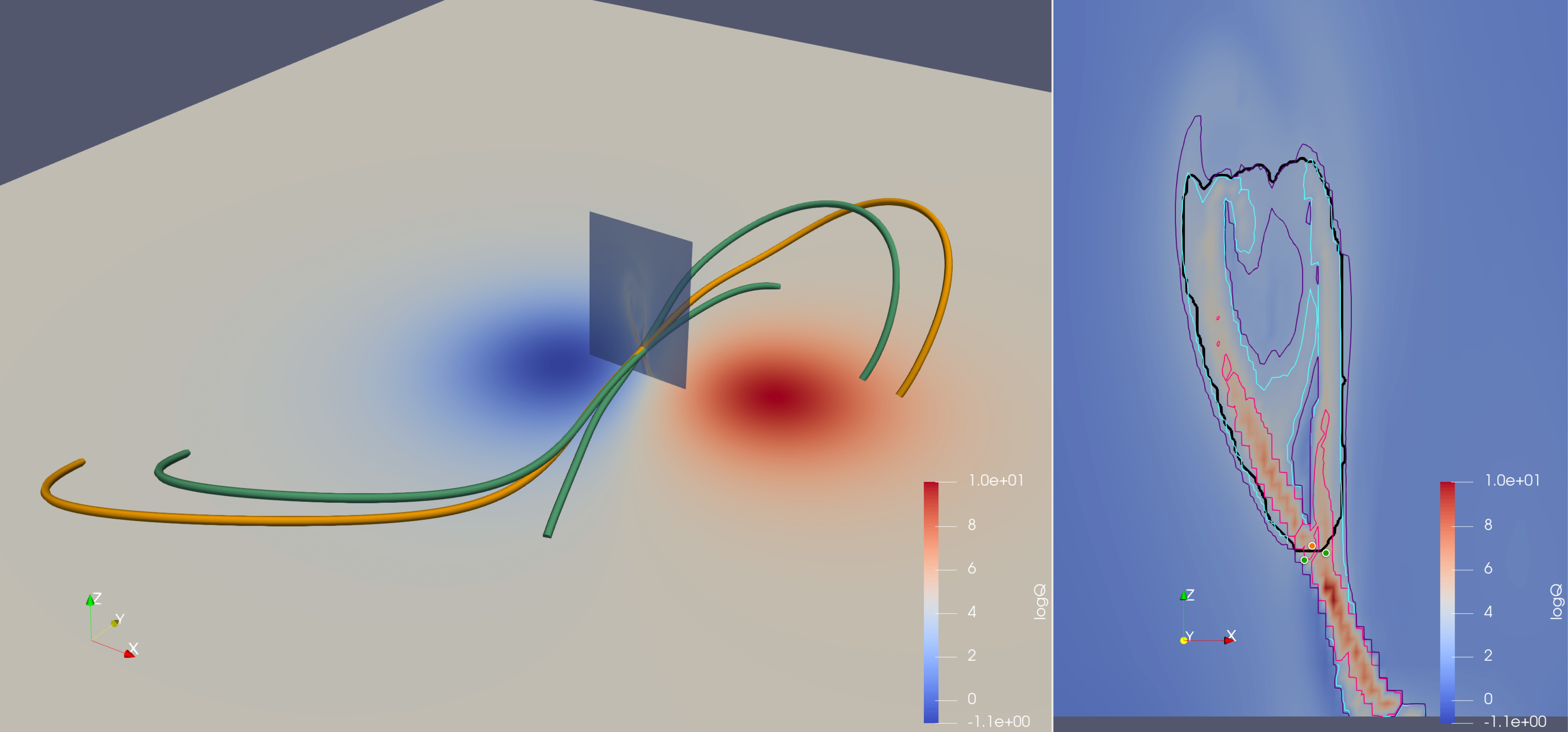}}
\caption{Left. 3D rendering of one MFR field line (orange) and two sheared arcades (green) at $t=100t_A$ that are separated by the MFR boundary as delineated by the black curve in the right panel. Right. Distribution of $\log Q$ overlaid with the contours of $\log Q = 2.7, 3.0, 5.0$ in purple, blue and magenta, respectively, at the plane y=-0.06 perpendicular to the MFR axis. The positions of the MFR field line and sheared arcades passing the plane are represented by the dots in orange and green, respectively.} \label{bottom_boundary}
\end{figure*}

\begin{figure*}
      \vspace*{-1.3\textwidth}
      \centerline{\hspace*{0.3\textwidth}
      \includegraphics[width=1.21\textwidth,clip=]{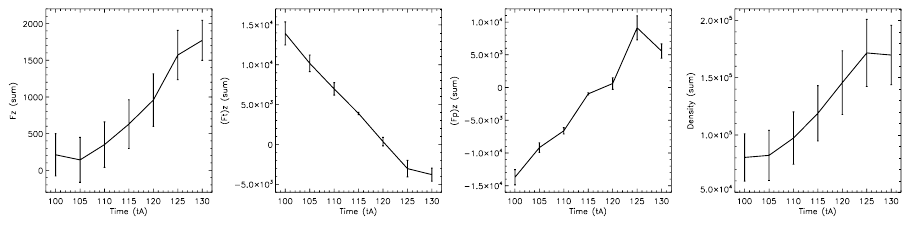}}\vspace*{-0.08\textwidth}
\caption{Temporal evolutions of the integrated vertical components of the Lorentz force, magnetic tension, magnetic pressure gradient and the mass density within the MFR cross section as shown in Figure \ref{f_force}a. The errors (vertical bars) are from the uncertainty in determining the MFR boundary.} \label{f_force1}
\end{figure*}

It is believed that the MFR boundary corresponds to the QSL, where the squashing degree $Q$ is very large. In practice, we identify the top and side boundaries of the MFR mainly by following the outer contours of $\log Q=3$. To enclose the bottom boundary of the MFR, the inner contours of $\log Q=5$, which clearly present the HFT configuration (Figure \ref{bottom_boundary}), are used for a reference. The integrated vertical components of the Lorentz force, magnetic tension force, magnetic pressure gradient and mass density are calculated by summing up the corresponding quantities within the boundary of the MFR at the plane. The acceleration caused by the Lorentz force and that by its two components are obtained by dividing the integrated forces by the integrated mass density. In order to estimate the uncertainty in determining the top and side boundaries of the MFR, we also use the contours of $\log Q=2.4$ and $\log Q=3.6$ instead of $\log Q=3.0$ to repeat the same procedure. The integrated quantities shown in Figure \ref{f_force1} and the accelerations in Figure \ref{f_force} are the averages of three measurements and their errors are corresponding standard deviations.

\restartappendixnumbering

\end{document}